# ORBIT FEEDBACK SYSTEM FOR THE STORAGE RING OF SRRC

C. H. Kuo, Jenny Chen, C. J. Wang, K. H. Hu, C. S. Chen, K. T. Hsu, SRRC, Hsinchu, Taiwan

Abstract

Orbit feedback system plays crucial roles for the operation of the 3$^{rd}$ generation light source. There are various issues in orbit feedback system should be addressed to achieve ultimate performance. The orbit feedback system in SRRC is upgraded recently to satisfy the requirement of demanding users. Based upon operational experiences of the last few years, new system was designed with more robustness and flexibility. Performance analysis tools are also developed to monitor system performance. Algorithms for feedback control, data acquisition and analysis are described and measurement is also presented.

## 1 INTRODUCTION

Orbit feedback system is used to eliminate orbit excursion due to various perturbation sources. Work to improve orbit stability start form 1995 with the orbit feedback system. This orbit feedback system has been applied incorporate with the insertion devices operation, include undulator (U5 and U9) and elliptical polarized undulator (EPU5.6). Suppress orbit drift and low frequency oscillation is also achieved. The orbit feedback system for the storage ring of SRRC is in upgrading to improve its performance. The effort include of increase feedback bandwidth, increasing sampling rate, compensate eddy current effect of vacuum chamber with filter, and enhance performance and robustness of the control rules. In this report, we will summary status of the orbit feedback development in SRRC.

## 2 EXISTING ORBIT FEEDBACK SYSTEM

A digital orbit feedback system [1,2] had been developed to suppress orbit disturbances caused by long-term drift, low-frequency oscillation and perturbation come from insertion devices operation. First, a linear response matrix is measured by taking electron beam position monitor (BPM) reading when the corrector is individually perturbed. Then, this response matrix is used to design a local orbit bump. The feedback controller is based on PID algorithm. Digital filtering techniques were used to remove noise of electron beam position reading, to compensate eddy current effect of vacuum chamber, and to increase bandwidth of orbit feedback loop. The infrastructure of digital orbit feedback system is composed of orbit acquisition system, gigabit fiber links, digital signal processing hardware and software, high precision digital-to-analog converters. From control points of view, The orbit feedback is a typical multiple input multiple output problems. The basic concept of the orbit feedback system is shown in the Figure 1. Technically, it is difficult to implement an analog matrix operation consisting of large amount of BPMs and correctors. Consequently, digital based feedback system is a natural way to implement orbit feedback system.

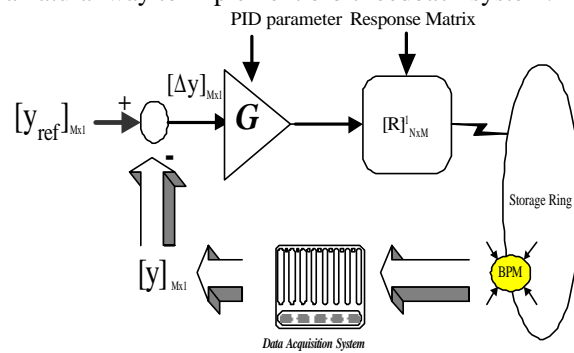

Figure 1. Basic concept of the orbit feedback system

There are two orbit feedback system in this control loop. One is local feedback, and another is global feedback. Integrating these two feedback loops for better coordination. System bandwidth of 10 - 100 Hz is necessary to suppress vibration and power supply ripple related beam motion, etc. The feedback system is integrated with the existed control system. BPMs data and correctors readback are updated into control system dynamic database in the period of 100 msec. Digital feedback system is bounded on I/O as well as computation. It is important to arrange the real time task and to arbitrate computer bus properly in order to optimize system performance.

### 2.1 Hardware structure

The hardware configuration of the corrector control system in SRRC is shown in Figure 2. The low layer is a VME crate system includes a PowerPC 604e CPU board and I/O interface cards. The front-end devices are connected to this system via analog and digital I/Os. A PowerPC based server system is used as the TFTP file server for downloads OS and mounted disk of network file server (NFS). All application programs are put on server disk. These programs are developed and debugged on client node to relief loading of server. The real-time multi-tasking kernel on the VME bus single board computer provided a satisfactory performance, reliability, and a rich set of system services. New device is easy to be created by that only modify device table file as if on line editing. The system can automatically boot and execute different applications in every VME node with the same operation system environments. The upload process handle device (analog input) and send acquisition data to database when receives the broadcast upload message from the Ethernet in every 0.1 second. The sampling rate of the feedback loop is 1 kHz by VME interrupt.

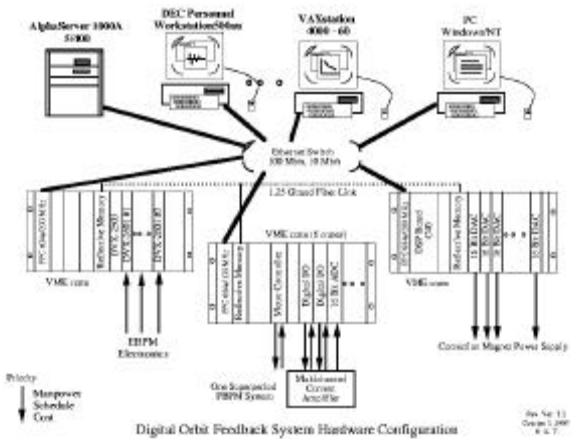

Figure 2. Hardware architecture of the orbit feedback system.

Present system consists of two VME crates, orbit server VME crate, corrector and DSP VME crate. Orbit server in charge to acquire BPM data every 1 msec and transfer the BPM data to corrector node by the aid of reflective memory. The host processor are also provide average BPM reading and update to control database every 100 msec, the average number is programmable. A JTAG emulator is used for the development of DSP program. The bus adapter is fit onto slot 1 of VME crate as system controller. Feedback software were developed and debugged on PC and downloaded to DSP board via front panel JTAG port. The DSP board carrying two TMS320C40 TIM modules handle all signal processing, including a digital low pass filter and PID controller. It takes 1 ms to complete feedback processes including data input, operation of PID, digital low pass filtering, matrix operation, BPMs data reading and corrector settings. The corrector setting had been applied to multi-channel 16 bit DACs with sub-μrad steering resolution. It is allowed remotely adjusted the corrector setting from graphical users interface during feedback loop is turn on.

Intrinsically, the performance of feedback system is limited by BPM and PBPM resolution. The PBPM data is directly acquired from multi-channel electrometer by A/D channel of VME crate, which are distributed in beam line. These crates are used as PBPM server node. It contain that PowerPC 604, reflective memory and 16 bit A/D card in each crate. The upper plate and low plate signals of PBPM are sent to PowerPC with VME bus. The vertical signals are sent to RM with PMC bus after transformation processing of two plates. This loop is synchronized with 1 ms PMC interrupt that is requested by server crate with RM and fiber link. There is 133 Mbytes per second data communication in PMC bus, so it is provided for PBPM data transfer quickly and largely. All data are collected to BPM server node. The orbit server provides fast beam position information to be used for feedback loop. It also provides slow orbit information for centralized database. The fast orbit information is sent to corrector via gigabit fiber linked reflective memory and computation needs in the VME crates.

## 2.2 Software structure of system

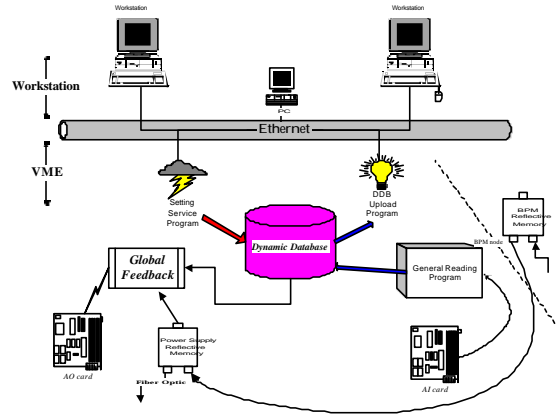

Figure 3: Software structure of the orbit feedback system.

The corrector node is supplied to corrector control with PowerPC, 16 bit D/A card and DSP card. The structure of application software is described in Figure 3. There are some service tasks in PowerPC. The setting task handles corrector setting when setting command arrived from database. It spawns child tasks to process all setting requests corresponding tasks to each incoming UPD facility setting packet. The reading process is triggered by the external 10 Hz clock from network when receives the broadcast upload message from Ethernet. And sends an event to wake up the data acquisition process. The data acquisition process is directly controller by remote login. The increased I/O card is easily updated by modified configure table file. All acquired data are broadcast to Ethernet in every 100 ms. The DDB process is server of shared memory that is coordinate the communication between reading process and setting process. It is also provided the data access from another process.

## 3 A EXAMPLE APPLICATION OF THE ORBIT FEEDBACK SYSTEM

Changing gap and phase parameters of insertion devices is essential in the operation scenario of a modern light source. Residue field of the insertion devices is the major perturbation source led to orbit excursion. It is hard to eliminate by lookup table compensation scheme. In routine operation of the storage ring, look-up table scheme is used to reduce orbit excursion and orbit feedback system is to keep the orbit change within micron level. During the gap change of U5 (4-meters undulator with 5 cm period), orbit change due to field error. The orbit changed without and with global orbit feedback while adjusting the U5 gap as indicated in Figure 4. The difference orbit is defined to be the orbit changed at 100 µm and 40mm from 219mm of U5 gap. The displacement of orbit was much smaller when the digital global feedback was turned on in comparison with the case when it was off.

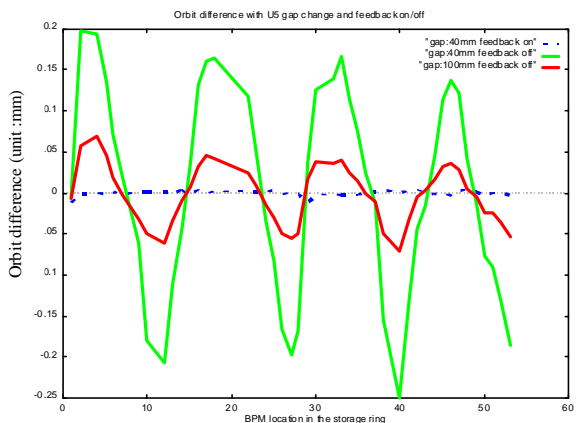

Figure 4: Orbit difference with U5 gap change and feedback on/off.

## 4 ORBIT FEEDBACK SYSTEM UPGRADE

Upgrading of the orbit feedback system is underway. It is planed that the new system will put into operation at early 2002. There is several reason of the upgrade: Firstly, to improve system maintainability and stability. The DSP board of the original feedback system is embedded in the corrector control VME crate. It is inconvenient for the development of the feedback system, interference between machine operation and feedback loop R&D is a troublesome problem. Secondly, the system is implemented in 1995 with slow DSP board; the functionality of the feedback loop is limited. There are not enough computing power to handle the increase demand of the number of BPMs and correctors. Selection of control algorithm is also limited. In new implementation, the DSP board is located on a separate VME crate. Ethernet based DSP development system was selected to provide remote access capability. The corrector node is loose coupled to the DSP VME crate. There are three VME crates in the upgraded system: BPM node, corrector node and DSP node. These three nodes are connected by reflective memory. Several of the fiber link reflective memory cards are tight together by one dedicated reflective hub that simplifies the wiring of the fiber link. The corrector node handles power-supply control. The DSP node handles graphic interface connection of feedback control, calculation of control algorithm and signal conversion from orbit information to correction of corrector. The correction value results are sent to corrector node and notice the host processor of corrector node by interrupt. The functional block diagram is shown in Figure 5. The planed photon BPM information is sent to DSP node by a private fast Ethernet network. Local data acquisition for PBPMs are planed to used compact PCI crate system. In the meantime, these data are sent to database of console level with control network.

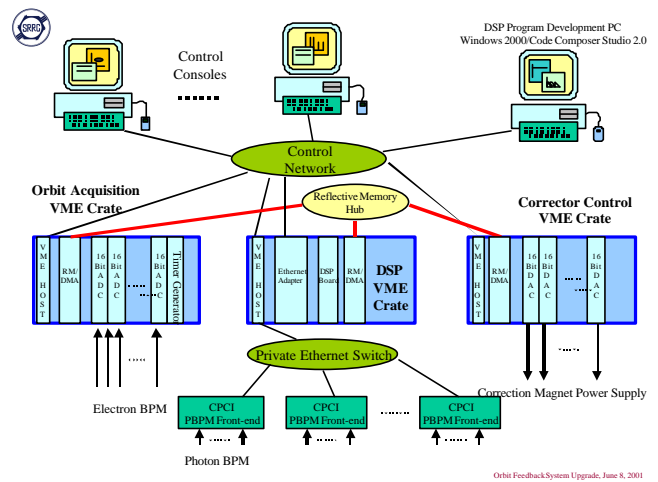

Figure 5: Functional block diagram of the upgrading orbit feedback system.

## 5 CONCLUSION

An orbit feedback system has been developed at SRRC in 1995. To satisfy the requirement of demanding user, the system is in upgrading now. The new system will provide sufficient computing power to execute various controls rules and improve the system maintainability at same time. The performance of this system will be improved as the hardware and software is upgraded and eliminated in any shortage of the existing system.